\documentclass{emulateapj} \usepackage{apjfonts}
\newcommand{\Ha}{$\rm{H} \alpha$}

\newcommand{\etal}{et~al.}
\newcommand{\NII}{\hbox{[{\rm N}\kern 0.1em{\sc ii}}]}
\newcommand{\HII}{\hbox{{\rm H}\kern 0.1em{\sc ii}}}
\newcommand{\kms}{\hbox{km~s$^{-1}$}}

\slugcomment{reubmitted March 9, 2015} 
\shorttitle{\sc ZFIRE: Mass-Metallicity Relation at z=2}
\shortauthors{\sc Kacprzak et~al.}

\begin{document}


\title{The Absence of an Environmental Dependence in the
  Mass-Metallicity Relation at z=2}


\author{\sc Glenn G. Kacprzak\altaffilmark{1}, Tiantian
  Yuan\altaffilmark{2}, Themiya Nanayakkara\altaffilmark{1}, Chiaki
  Kobayashi\altaffilmark{2,3}, Kim-Vy H. Tran\altaffilmark{4}, Lisa
  J. Kewley\altaffilmark{2}, Karl Glazebrook\altaffilmark{1},
  Lee Spitler\altaffilmark{5,6}, Philip Taylor\altaffilmark{3}, Michael
  Cowley\altaffilmark{5,6}, Ivo Labb\'e\altaffilmark{7}, Caroline
  Straatman\altaffilmark{7}, Adam Tomczak\altaffilmark{4}}

\altaffiltext{1}{Swinburne University of Technology, Victoria 3122,
Australia {\tt gkacprzak@astro.swin.edu.au}}
\altaffiltext{2}{Research School of Astronomy and Astrophysics, The Australian National University, Cotter Road, Weston Creek, ACT 2611, Australia}
\altaffiltext{3}{Centre for Astrophysics Research, Science and Technology Research Institute, University of Hertfordshire, AL10 9AB, UK}
\altaffiltext{4}{George P. and Cynthia Woods Mitchell Institute for Fundamental Physics and Astronomy, and Department of Physics and Astronomy, Texas A\&M University, College Station, TX 77843-4242, USA}
\altaffiltext{5}{Department of Physics and Astronomy, Macquarie University, Sydney, NSW 2109, Australia}
\altaffiltext{6}{Australian Astronomical Observatories, PO Box 915 North Ryde NSW 1670, Australia}
\altaffiltext{7}{Leiden Observatory, Leiden University, P.O. Box 9513, 2300 RA Leiden, The Netherlands}

\begin{abstract}

We investigate the environmental dependence of the mass-metallicity
relation at $z=2$ with MOSFIRE/Keck as part of the ZFIRE survey.
Here, we present the chemical abundance of a Virgo-like progenitor at
$z=2.095$ that has an established red sequence.  We identified 43
cluster ($<z>=2.095\pm0.004$) and 74 field galaxies
($<z>=2.195\pm0.083$) for which we can measure metallicities. For the
first time, we show that there is no discernible difference between
the mass-metallicity relation of field and cluster galaxies to within
0.02dex. Both our field and cluster galaxy mass-metallicity relations
are consistent with recent field galaxy studies at $z\sim2$. We
present hydrodynamical simulations for which we derive mass-metallicity
relations for field and cluster galaxies. We find at most a 0.1dex
offset towards more metal-rich simulated cluster galaxies.  Our
results from both simulations and observations are suggestive that
environmental effects, if present, are small and are secondary to the
ongoing inflow and outflow processes that are governed by galaxy halo
mass.

\end{abstract}




\section{Introduction}
\label{sec:intro}

%%%%%%%%%%%%%%%%%%%%%%%%%%%%%%%%%%%%%%%%%%%%%%%%%%%%%%%%%%%%%%%%%%

\begin{figure*}
\begin{center}
\includegraphics[angle=0,scale=1.42]{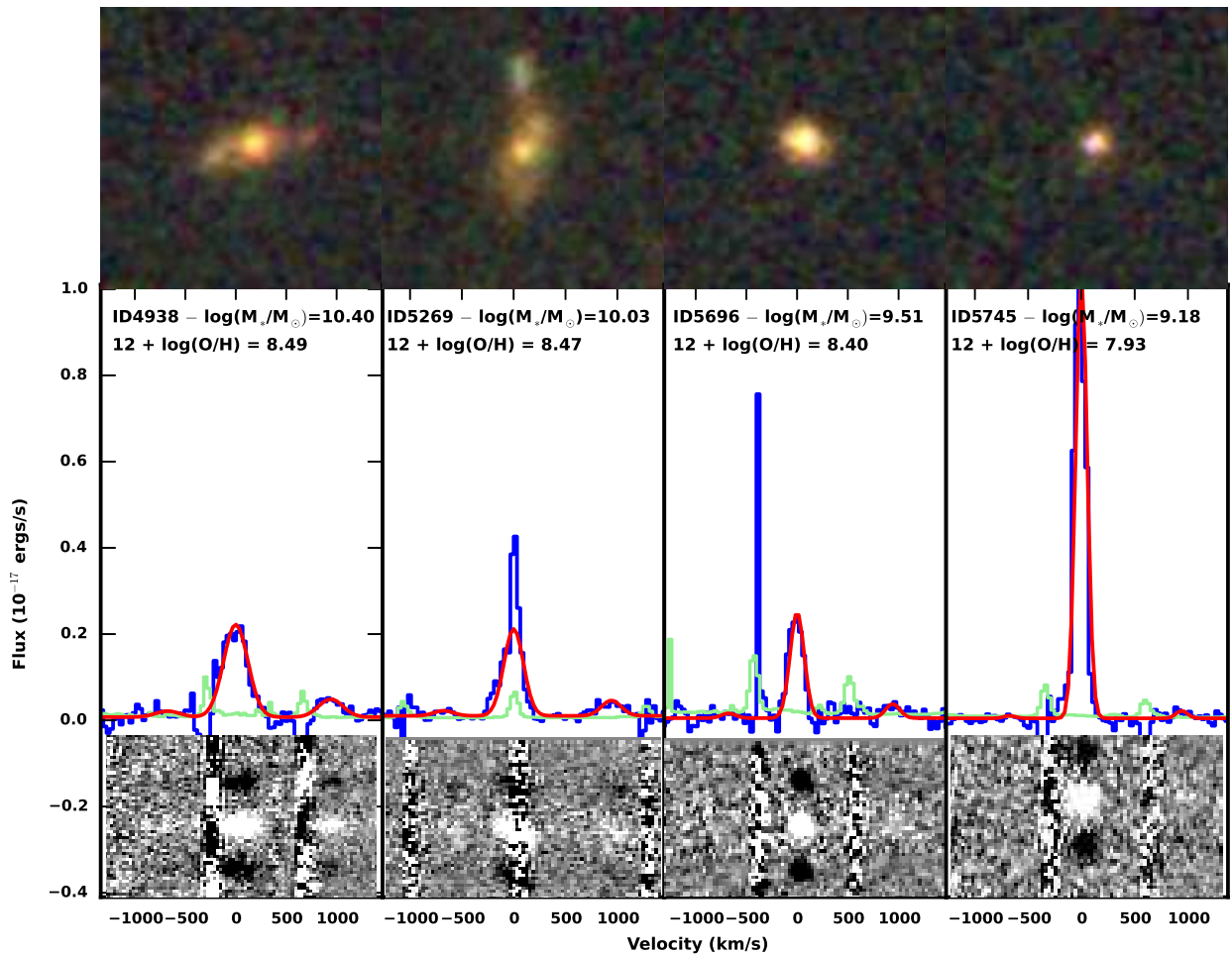}
\caption[angle=0]{Representative sample of flux-calibrated 1D and 2D
  MOSFIRE spectra of four cluster members in order of decreasing
  stellar mass. The spectra are plotted with respect to the systemic
  velocity of each galaxy, as definded by their nebular
  emission-lines. Both the {\NII} doublet and {\Ha} are shown in the
  1D and 2D spectra. Strong sky-lines are seen as long virtical lines
  in the 2D spectra. The 1D data (blue), 1~$\sigma$ error spectra
  (green) and best-fit Gaussian models (red) are shown. The
  metallicity for each object is also listed. The top panels show
  $4''\times 4''$ three color {\it HST} images, using the F814W, F125W
  and F160W filters, obtained from the publicly available CANDELS data
  \citep{grogin11,koekemoer11,skelton14}. }
\label{fig:MOS}
\end{center}
\end{figure*}

%%%%%%%%%%%%%%%%%%%%%%%%%%%%%%%%%%%%%%%%%%%%%%%%%%%%%%%%%%%%%%%%%%%

The evolution of galaxies is likely dependent on halo mass and
regulated by baryonic feedback cycles \citep[e.g.,][]{dekel09}. In
addition, environment plays a role in removing and truncating gas
reservoirs while increasing the frequency of galaxy
interactions/mergers, which could cause differences in morphology and
color seen locally for field and cluster galaxies
\citep[e.g.,][]{dressler80}. However, selecting low redshift galaxies
by their stellar ages and stellar masses shows that their observable
properties are independent of environment \citep[see][for a
review]{blanton09}. This is also confirmed by studies showing that the
mass-metallicity relation of local field and cluster star-forming
galaxies are consistent
\citep{mouchine07,ellison09,scudder12,huges13}. If there is a
difference in the mass-metallicity relation as a function of
environment at $z\sim0$, then it is observationally constrained to be
around 0.05dex \citep{ellison09} while simulations constrain it to be
less than $\sim$0.05dex \citep{dave11}.

Although it is clear that a mass-metallicity relation exists at
$z\sim2$ \citep{erb06,sanders14,steidel14}, it is unclear if it
differs as a function of environment at high redshift where clusters
are beginning to form.  Exploring the environmental influence on
galaxies at high redshift is difficult since only a handful of
spectroscopically confirmed clusters with $z\gtrsim2$ exist
\citep[e.g.,][]{kurk04,galametz13,gobat2013,shimakawa14,valentino14,yuan14}. Only
three studies have attempted to address the effects of environment on
the mass-metallicity relation using small samples of galaxies, or
inhomgenous field and cluster selections, and have produced
conflicting results. A star-forming protocluster containing 6 galaxies
at $z=1.99$ was found to be up to 0.25dex poorer in metals than their
field counterparts \citep{valentino14}, while two star-forming
protoclusters at $z=2.2$ (24 galaxies) and at $z=2.5$ (36 galaxies)
were found to be more enriched than those of field galaxies for
M$_\star<10^{11}$~M$_{\odot}$ \citep{shimakawa14}. On the other-hand,
\citet{kulas13} found for 20 star-forming protocluster galaxies that
they do not exhibit a dependence of metallicity on mass (i.e., zero
slope) with the low-mass protocluster galaxies showing an enhancement
in metallicity compared to field galaxies. Although there appears to
be some disagreement between studies, these $\sim$0.15dex differences
are marginally detected at less than two sigma.

Here, we present Keck/MOSFIRE observations from the ZFIRE
survey\footnote{http://zfire.tamu.edu} (T. Nanayakkara et al. in prep)
in order to clarify the debate on the dependence between the
mass-metallicity relation and environment found at $z\sim2$.  We
identified 43 cluster ($<z>=2.095\pm0.004$) and 74 field galaxies
($<z>=2.195\pm0.083$) \citep{spitler12,yuan14} for which we can
measure metallicities.  This study is unique given the large number of
galaxies obtained for a single cluster, with field galaxies selected
the same way and observed simultaneously with the same instrument.  In
\S~\ref{sec:data} we describe the data and our metallicity
measurements.  In \S~\ref{sec:results}, we present the
mass-metallicity relation for field and cluster galaxies at
$z\sim2$. Our data are consistent with previously published field
galaxy relations, however contrary to other cluster galaxy works, we
do not find a difference in the relation as a function of
environment. We further show that this result is consistent with
current simulation predictions. We end with concluding remarks in
\S~\ref{sec:conclusion}.

Throughout, we adopt a $h=0.70$, $\Omega_{\rm M}=0.3$,
$\Omega_{\Lambda}=0.7$ cosmology.

%%%%%%%%%%%%%%%%%%%%%%%%%%%%%%%%%%%%%%%%%%%%%%%%%%%%%%%%%%%%%%%%%%

\begin{figure*}
\begin{center}
\epsscale{1.15}\plottwo{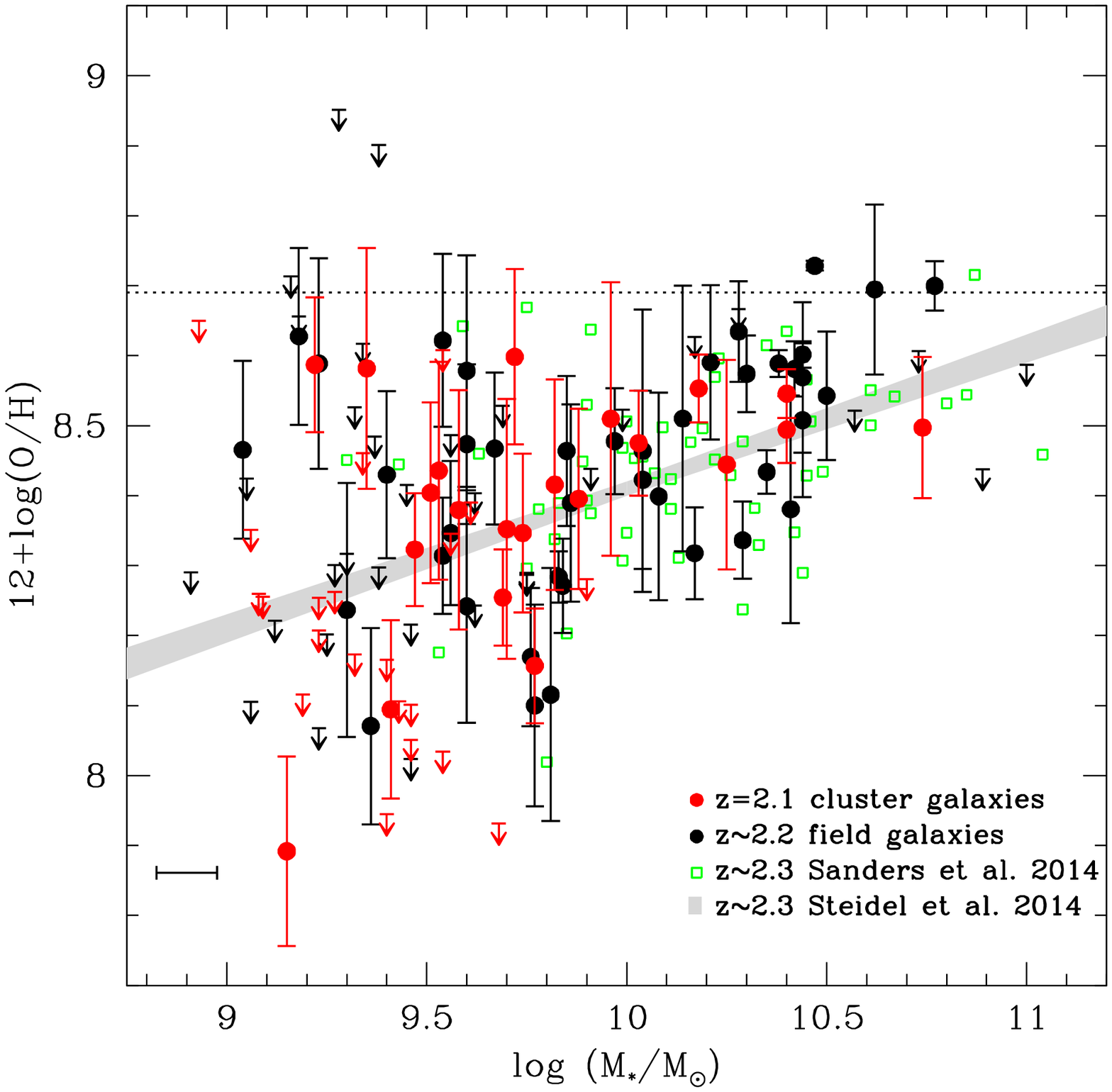}{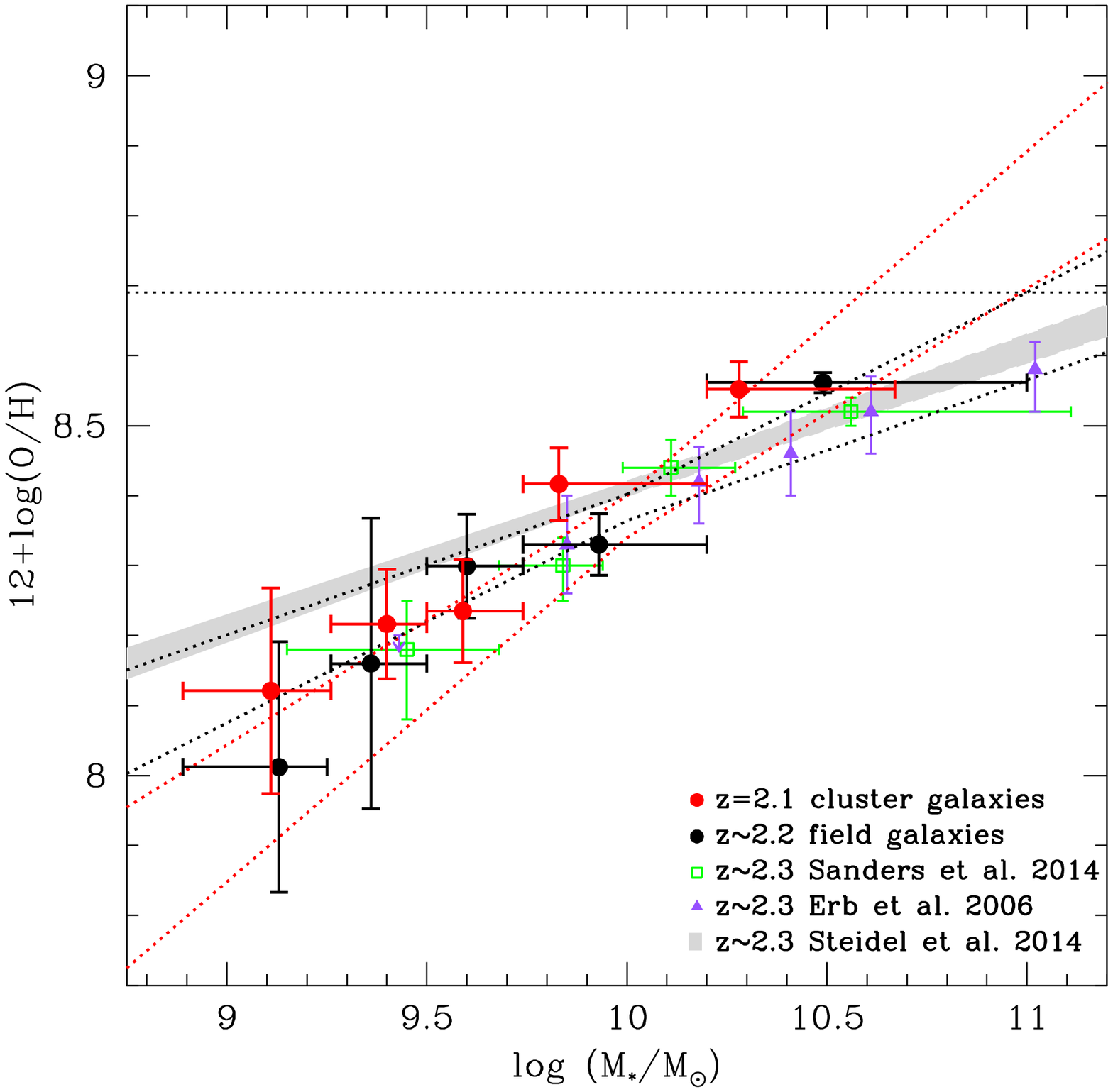}
\caption{ (left) The mass-metallicity relation for star-forming
  $z=2.1$ cluster (red) and $z\sim 2.1$ field galaxies
  (black). Circles indicate metallicity measurements where both {\Ha}
  and \NII are detected with greater than 3$\sigma$ significance and
  downward arrows are quoted as 1$\sigma$ limits on \NII. The average
  error in determining the mass of galaxies using ZFOURGE photometry
  and FAST of $\pm$0.076 is shown in the bottom left. The green
  circles show the individual MOSDEF detections \citep{sanders14},
  while the grey shaded region shows the fit to KBSS-MOSFIRE
  \citep{steidel14}. The dotted horizontal line solar abundance
  \citep{asplund09}. --- (right) The dotted lines are bootstrap fits,
  with 1$\sigma$ limits, to the mass-metallicity data from the left
  panel for cluster (red) and field (black) galaxies (see text for
  details). The solid points are stacked spectra within a given mass
  bin and note that these are consistent with the fitted data. We
  further show additional binned data from the literature. Note both
  cluster and field galaxies are consistent within the 1$\sigma$
  errors.}
\label{fig:MMR}
\end{center}
\end{figure*}
%%%%%%%%%%%%%%%%%%%%%%%%%%%%%%%%%%%%%%%%%%%%%%%%%%%%%%%%%%%%%%%%%%

\section{MOSFIRE Spectroscopic Observations and Sample}
\label{sec:data}

The $z=2.095$ cluster was previously identified using the photometric
redshifts \citep{spitler12} from the medium-band FourStar Galaxy
Evolution Survey (ZFOURGE, Straatman et al. in prep).  As part of the
spectroscopic followup to ZFOURGE, the ZFIRE survey used Keck/MOSFIRE
to spectroscopically confirm the existence of a Virgo-like progenitor
at $z=2.095$ containing at least 57 members with a velocity dispersion
of 550~{\kms} \citep{yuan14}.  Furthermore, the ZFIRE survey yielded
an additional 123 spectroscopic redshifts of field galaxies with
$1.98\leq z\leq3.26$. The spectroscopic targets were selected using
the photometric redshifts from the K-band selected catalog from
ZFOURGE \citep[][Straatman et al. in prep]{spitler14}, which have an accuracy of $\Delta
z/(1+z_{\hbox{spec}})=2\%$ \citep{tomczak14,yuan14}.

The MOSFIRE near-infrared K-- and H--band spectroscopic observations,
data reduction and flux calibration procedures are described in
\citet{yuan14} with additional details to be presented T. Nanayakkara
et al. (in prep). All spectra are calibrated to vacuum wavelengths.
Our typical 3$\sigma$ flux limit is
1.8$\times$10$^{-18}$~ergs/s/cm$^2$.

Gaussian profiles were simultaneously fit to {\Ha} and {\NII}
emission-lines to determine their total flux. The line centers and
velocity widths were tied together for a given pair of lines.
Examples of our 2D and 1D MOSFIRE spectra and line-fits are shown in
Figure~\ref{fig:MOS} for galaxies that have stellar masses ranging
between log(M/M$\odot$)=10.4--9.2 (also see \citet{yuan14} for
additional examples).

We compute a gas-phase oxygen abundance for each galaxy using the N2
relation of \citet{pettini04} where 12+log(O/H)=8.90+0.57$\times$N2
(N2$\equiv$log(\NII/{\Ha})). The \citet{pettini04} relation is
established from a sample of low-redshift extragalatic {\HII} regions
for which both N2 and oxygen abundances are directly
measured. Applying this relation to high redshift galaxies implies the
caveat/assumption that the ionization parameters and electron
densities are similar to those of low redshift galaxies, however, it
is known that typical ionization parameters are much higher for
$z\sim2$ galaxies with log($U$)$\geq$7.3
\citep[e.g.,][]{shirazi14,steidel14,kewley15}.  It is important to
note however, that \citet{kewley15} has shown that our $z\sim2$ field
and cluster galaxies have consistent ISM conditions, which allows for
a meaningful direct comparison between the mass-metallicity relations
of field and cluster galaxies.

From the parent sample of \citet{yuan14}, we identify 49 cluster and
86 field galaxies with an {\Ha} flux detected at greater than
3$\sigma$ significance.  Using the ZFOURGE AGN catalog (M. Cowley et
al. in prep), we remove AGNs from our sample (7 from the field, 3 from
the cluster). An AGN is identified as, or a combination of, an
infrared source \citep[following methods of][]{donley12}, an X-ray
source \citep[following methods of][]{szokoly04} or a radio source
(following methods of G.~Rees et al. 2015, in prep).  In addition, we
further require log(\NII/{\Ha})$< -0.3$ \citep{sanders14}, which
removes 8 additional AGN (5 field, 3 cluster).

Our final sample contains 43 cluster ($<z>=2.095\pm0.004$) and 74
field galaxies (1.98$\leq z\leq$2.56, $<z>=2.195\pm0.083$) for which
we can measure metallicities.  We require a 3$\sigma$ detection
significance level for {\NII}, otherwise 1$\sigma$ detection limits
are shown. Our final sample contains 22 metallicity measurements and
21 limits for the cluster galaxies and 41 metallicity measurements and
33 limits for the field galaxies.

We use stellar masses computed from the ZFOURGE photometry using
\citet{bruzual03} stellar population models with FAST \citep{kriek09},
assuming exponentially declining star formation histories, solar
metallicity, a \citet{chabrier03} initial mass function and
constrained to the spectroscopic redshift \citep[see][for
  details]{tomczak14}.

2D Kolmogorov-Smirnov test shows that the cluster and field galaxies
have statistically consistent (P(KS)$<1\sigma$) observational
properties such as extinction, (determined by FAST) and star-formation
rates (determined by H$\alpha$ -- to be discussed in Kacprzak et al.,
in prep).

\section{Results}\label{sec:results}

\subsection{ZFIRE Observations}

In Figure~\ref{fig:MMR} (left), we show the mass-metallicity relation
for our field (black) and cluster (red) galaxies at $z\sim2$,
including 1$\sigma$ limits. We have metallicity measurements for a
significant range of galaxy masses from
$8.9\leq$log(M/M$_{\odot}$)$\leq11.0$. We further show the 53
individual detection from MOSDEF \citep{sanders14} and the fitted
relation from KBSS-MOSFIRE \citep{steidel14}. Note that the scatter in
the data is similar to the scatter in the MOSDEF distribution and our
data seem to follow the fitted data from KBSS-MOSFIRE. Furthermore,
our cluster and field galaxies appear to have similar mass-metallicity
distributions.

In Figure~\ref{fig:MMR} (right), we show a bootstrap fit (1000 times)
to the data from Figure~\ref{fig:MMR} (left) using 12+log(O/H)$=y_i
+m_i(M-10)$.  We fit the relation including the 1$\sigma$ limits using
the expectation-maximization maximum-likelihood method of
\citet{wolynetz79}. Including limits we find for the cluster
$y_{c}=8.370\pm0.030$ and $m_{c}=0.424\pm0.068$ while for the field we
find $y_{f}=8.384\pm0.018$ and $m_{f}=0.245\pm0.044$. These fits,
along with their 1$\sigma$ errors, are shown as dotted lines in
Figure~\ref{fig:MMR} (right).  From the fitted data, we find that both
cluster and field galaxies exhibit similar mass-metallicity relations
and are consistent within 1$\sigma$.  The maximum separation between
the fitted field and cluster galaxy metallicity is at most
0.014$\pm$0.035dex. There could be a difference in the slope between
field and cluster galaxies, which is mostly driven by the larger
number of low mass cluster metallicity limits.

To explore the similarities between the field and cluster galaxy data,
we further stack the spectra in five mass bins with roughly equal
numbers of galaxies per bin. We stacked the spectra, weighting by the
uncertainty spectrum, to determine the typical metallicity in a given
mass bin shown in Figure~\ref{fig:MMR}.  We found no discernible
difference if we weighted each spectrum by its {\Ha} luminosity.
Again, the stacked spectra show that cluster and field galaxies have
equivalent mass-metallicity relations and are consistent within the
1$\sigma$ errors. The stacked spectra are also consistent with the
fits to the indvidual datapoints.

The fits and stacked data for both our field and cluster galaxies are
consistent with the previous results of MOSDEF \citep{sanders14},
KBSS-MOSFIRE \citep{steidel14} and \citet{erb06}. However, the low
mass end of the field galaxies marginally deviate away from the
KBSS-MOSFIRE fit, but is consistent with their binned data (not
shown). Any observed minor differences could due to sample selection
biases (KBSS-MOSFIRE is rest-frame ultraviolet selected while MOSDEF
and ZFIRE are rest-frame optical selected) or how detection
significance levels and fits were conducted. However, given that our
cluster and field data are consistent with all previous works, further
validates the lack of a difference between field and cluster galaxies.

%%%%%%%%%%%%%%%%%%%%%%%%%%%%%%%%%

\subsection{Cosmological Simulations}

To further examine if the environment of galaxies actually influences
their metallicities, we perform a similar analysis using cosmological
simulations. The simulations are performed with a Gadget-3 based
hydrodynamical code that includes important baryon physics such as
star formation, feedback from supernovae and AGN, and chemical
enrichment from Type~II and Ia supernovae, and asymptotic giant branch
stars with Kroupa IMF \citep[see][for the
  details]{kobayashi07,taylor14}. Metallicities derived for Kroupa IMF
are virtually identical for to those derived for a Chabrier IMF. The
input nucleosynthesis yields are in excellent agreement with the
observed elemental abundances in the Milky Way Galaxy from carbon to
zinc \citep{kobayashi11}.  

Different from the \cite{dave11} simulations, we include AGN feedback,
whereby AGN-driven winds eject metals from massive galaxies to the
intergalactic medium, and some less-massive galaxies undergo external
enrichment depending on the environment.  However we find that for the
mass-metallicity relation, the effect of AGN is small and the trend
originate mainly from supernova/hypernova-driven galactic winds, which
is much larger in less-massive galaxies \citep{kobayashi07}.

The simulations are run for a 25$h^{-1}$~Mpc box with two different
initial conditions. The cluster simulation is chosen from ten
realization to have a strongest central concentration, which gives the
most massive galaxy with the stellar mass $\sim 10^{12}M_\odot$
\citep{taylor14} at $z=0$, while the field simulation is same as used
in \citet{kobayashi07} and does not contain any central
concentrations.  The stellar masses are estimated by fitting a
core-S{\'e}rsic profile, and the oxygen abundances are measured in
15$h^{-1}$~kpc weighted by the star formation rates of gas particles
to be comparable to our emission-line observations.  At $z=0$, the
mass-metallicity relations of simulated galaxies are in good agreement
with the observed mass-metallicity relations both for stellar and
gas-phase metallicities \citep{taylor15}.  These relations evolve as a
function of time, with lower metallicities and a steeper slope at
higher redshifts (Taylor \& Kobayashi 2015, in prep.).

%%%%%%%%%%%%%%%%%%%%%%%%%%%%%%%%%%%%%%%%%%%%%%%%%%%%%%%%%%%%%%%%%%

\begin{figure}
\begin{center}
\includegraphics[angle=0,scale=0.44]{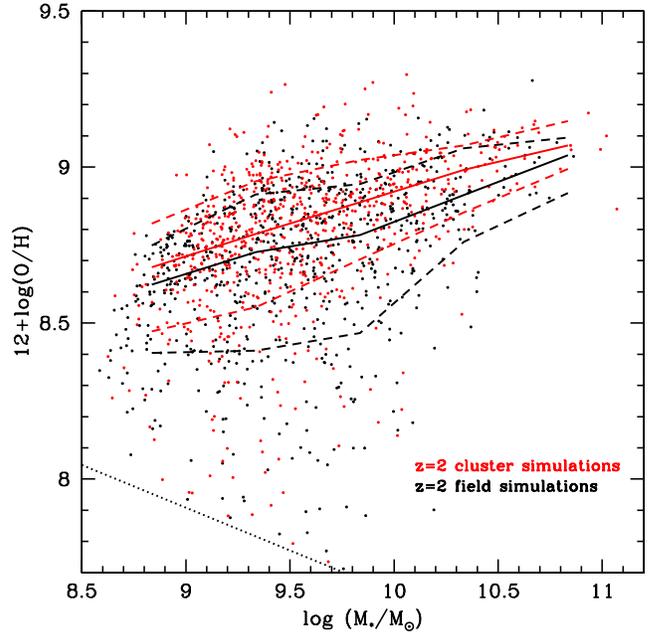}%../lk3.eps}
\caption[angle=0]{Mass-metallicity relation for simulated field
  (black) and cluster (red) galaxies. The solid and dashed lines
  show the moving median with 68\% of confidence level. The dotted
  line is an observational limit of our ZFIRE survey (see text for
  details). We find that both field and cluster galaxies have
  consistent mass-metallicity relations with maximum separations of
  0.1dex.}
\label{fig:simMMR}
\end{center}
\end{figure}
%%%%%%%%%%%%%%%%%%%%%%%%%%%%%%%%%%%%%%%%%%%%%%%%%%%%%%%%%%%%%%%%%%

Figure~\ref{fig:simMMR} shows the mass-metallicity relations of the
``cluster'' and ``field'' simulations at $z=2$. We apply an
observationally-based detection limit on the simulated
mass-metallicity distribution.  The N2-based metallicity measurement
is confined by the detection limit of {\NII}. Based on a few simple
assumptions, \citet{yuan13} used the SFR-mass relation, flux detection
limit and redshift to determine observational detection limit on the
mass-metallicity relation. For the ZFIRE survey, our typical 3$\sigma$
flux limit is 1.8$\times$10$^{-18}$~ergs/s/cm$^2$ resulting in the
limiting relation of:
12+log(O/H)~>~$-0.274\times$log(M$_{\star}$/M$_{\odot}$)+10.371.  This
observational detection limit is important for understanding the
incompleteness and biases due to observational limitations. While all
of our observational data reside above this limit, we apply it to our
simulated data (dotted line).

In Figure~\ref{fig:simMMR}, the solid and dashed lines show the moving
median with 68\% confidence levels also shown.  Both the field and
cluster galaxies exhibit similar mass-metallicity relationships within
1$\sigma$. The maximum separation between the fitted field and cluster
galaxy metallicity is at most 0.1dex. This is consistent with previous
results from $z=0$ simulations that constrain the difference to be
less than $\sim$0.05dex \citep{dave11}.

Note that there is an $\sim0.5$dex offset in the zero-point between
simulations and observations, which can be partly attributed to the
uncertain normalization in metallicity measurements
\citep[e.g.,][]{kewley08,dave11}.

\section{Conclusions}\label{sec:conclusion}

High redshift protoclusters are ideal laboratories to study the
possible influences of environment on galaxies. It is important to
establish whether the environment can affect chemical evolution via
restricting inflow and outflow, which may be responsible for galaxies
migrating from the blue to the red sequence, or if a galaxys' chemical
evolution is established early and is solely dependent on its internal
evolutionary processes.

The environment may already be affecting the morphological properties
of galaxies at $z=2$. In our cluster, quiescent galaxies have similar
colors and sizes relative to field quiescent galaxies, however,
cluster star-forming galaxies are larger and redder on average when
compared to field star-forming galaxies -- suggesting environment can
transform some galaxy properties at this early stage of formation
\citep{allen15}.

It is a concern when searching for metallicity differences as a
function of environment that the precise metallicities derived for high
redshift galaxies are likely incorrect due to their different ISM
conditions relative to $z=0$ galaxies -- where emission-line
metallicity indicators are calibrated
\citep[e.g.,][]{pettini04}. However, this does not affect the relative
offsets between the field and cluster galaxies given \citet{kewley15}
has shown that for our cluster, both field and cluster galaxies have
similar ISM conditions.

Here, we present the chemical abundance of a Virgo-like progenitor at
$z=2.1$ \citep{yuan14} that has an established red sequence
\citep{spitler12}. Using our sample of field galaxies at a similar
redshift, we show that the mass-metallicity relation of field and
cluster galaxies are consistent with 1$\sigma$ errors.  We further
show that both our field and cluster galaxies have consistent
mass-metallicity relations when compared to other field galaxy surveys
at $z\sim2$ \citep{erb06,sanders14,steidel14}. Although we cannot rule
out weak environmental trends in chemical enrichment, our analysis
shows that the difference between field and cluster galaxies in the
mass-metallicity relation is less than 0.02dex.

Our simulations have shown that cluster galaxies may be marginally
more metal rich than field galaxies by at most 0.1dex (although
consistent within the scatter). This could be due to the removal of
gas from the outskirts of galaxies which can produce an observed
increase in metallicity by $\sim$0.1dex \citep{huges13} or due to the
shorter gas recycling times in denser environments
\citep{oppenheimer08}.  However, given that the offset is small, it is
suggestive that these effects do not play a significant role in the
chemical evolution of galaxies, especially given that there is no
observed differences in mass-metallicity relation of field and cluster
galaxies at $z=0$ \citep{mouchine07,scudder12,huges13}.

Our results from the simulations and observations are suggestive that
environmental effects, if present, are secondary to the ongoing
internal processes within $z\sim 2$ galaxies that are likely governed
by halo mass.

%%%%%%%%%%%%%%%%%%%%%%%%%%%%%%%%%%%%%%%% 

\acknowledgments  

We thank Ryan Sanders and Alice Shapley for providing MOSDEF survey
data.  We thank Jackson Cunningham for his contributions during this
initial stages of this project. GGK was supported by an Australian
Research Council Future Fellowship FT140100933. C. K. thanks the RSAA
distinguished visitor program. KG acknowledges support from
DP130101460 and DP130101667. Data was obtained at the W.M. Keck
Observatory, which is operated as a scientific partnership among the
California Institute of Technology, the University of California and
the National Aeronautics and Space Administration. The Observatory was
made possible by the generous financial support of the W.M. Keck
Foundation. Observations were supported by Swinburne Keck programs
2013B\_W160M and 2014A\_W168M and ANU Keck programs 20132B\_WZ295M and
2014A\_Z225M. Part of this work was supported by a NASA Keck PI Data
Award, administered by the NASA Exoplanet Science Institute. The
authors wish to recognize and acknowledge the very significant
cultural role and reverence that the summit of Mauna Kea has always
had within the indigenous Hawaiian community. We are most fortunate to
have the opportunity to conduct observations from this mountain.

%% Included in this acknowledgments section are examples of the
%% AASTeX hypertext markup commands. Use \url without the optional [HREF]
%% argument when you want to print the url directly in the text. Otherwise,
%% use either \url or \anchor, with the HREF as the first argument and the
%% text to be printed in the second.

%doing the math in section~\ref{bozomath}.
%More information on the AASTeX macros package is available \\ at
%\url{http://www.aas.org/publications/aastex}.
%For technical support, please write to
%\email{aastex-help@aas.org}.


%% in parentheses, after the keyword, but they will not be verified.

{\it Facilities:} \facility{Keck I (MOSFIRE)}.

\end{document}